\documentclass[12pt,preprint]{aastex}

\def\grb{GRB\,020405}

\begin{document}

\title{\large A Radio Flare from \grb: Evidence for a Uniform Medium
Around a Massive Stellar Progenitor}

\author{E. Berger\altaffilmark{1} and A. M. Soderberg\altaffilmark{1}}
\affil{Division of Physics, Mathematics, and Astronomy,
California Institute of Technology 105-24, Pasadena, CA 91125}
\email{ejb@astro.caltech.edu; ams@astro.caltech.edu}
\author{D.~A. Frail\altaffilmark{2}}
\affil{National Radio Astronomy Observatory, Socorro, NM 87801}
\email{dfrail@nrao.edu}
\author{S. R. Kulkarni\altaffilmark{1}}
\affil{Division of Physics, Mathematics, and Astronomy,
California Institute of Technology 105-24, Pasadena, CA 91125}
\email{srk@astro.caltech.edu}

\begin{abstract}

We present radio observations of \grb{} starting 1.2 days after the 
burst, which reveal a rapidly-fading ``radio flare''.
Based on its temporal and spectral properties, we interpret the radio
flare as emission from the reverse shock.  This scenario rules out a
circumburst medium with a radial density profile $\rho\propto r^{-2}$
expected around a mass-losing massive star, since in that case the
reverse shock emission decays on the timescale of the burst duration
$t\sim 10^2$ s.  Using published optical and X-ray data, along with
the radio data presented here, we further show that a self-consistent
model requires collimated ejecta with an opening angle, $\theta_j\sim
6^\circ$ ($t_j\approx 0.95$ day).  As a consequence of the early jet
break, the late-time ($t>10$ day) emission measured with the {\it
Hubble Space Telescope} significantly deviates from an extrapolation
of the early, ground-based data.  This, along with an unusually red
spectrum, $F_\nu\propto \nu^{-3.9}$, strengthens the case for a
supernova that exploded at about the same time as \grb{}, thus
pointing to a massive stellar progenitor for this burst.  This is the
first clear association of a massive progenitor with a uniform
medium, indicating that a $\rho\propto r^{-2}$ profile is not a
required signature, and in fact may not be present on the lengthscales
probed by the afterglow in the majority of bursts.
\end{abstract}

\keywords{gamma-rays:bursts --- stars:mass loss}

\section{Introduction}
\label{sec:intro}

Over the past few years several indirect lines of evidence have
emerged in favor of massive stars as the progenitors of long-duration
gamma-ray bursts (GRBs).  These include the distribution of offsets of
GRBs from their host centers \citep{bkd02}, the absence of optical
afterglows from dark GRBs \citep{dfk+01,pfg+02}, high column densities
toward several GRBs \citep{gw01}, and the inference of very high star
formation rates in several GRB host galaxies
\citep{bkf01,fbm+02,bck+02}.

Perhaps the most convincing evidence comes from the detection
of late-time ($t\sim 20$ days) red bumps dominating the optical
emission from several afterglows (e.g.~\citealt{bkd+99}).  While
several interpretations of these bumps have been suggested 
\citep{eb00,wd00,rdm+01,rei01}, the preponderance of spectral and
temporal evidence (e.g.~\citealt{bkp+02,gsw+02}) indicates that these
bumps are due to radioactive Nickel emission from supernovae (SNe)
accompanying the bursts.  These observations lend support
to the collapsar model \citep{woo93,mw99}, in which the core of a
massive star collapses to a black hole, which then accretes matter and
powers the GRB, while the rest of the star produces a supernova.

A seemingly unavoidable consequence of this scenario is that the GRB
ejecta should expand into a medium modified by mass loss from the
progenitor star.  To first order, the expected density profile is
$\rho\propto r^{-2}$, arising from constant mass loss rate and wind
velocity.  Extensive efforts have been made to find evidence for such
a density profile based on broad-band observations of the afterglow
emission (e.g.~\citealt{cl00,bsf+00,pk02}).  Unfortunately, these
studies have been inconclusive in distinguishing between a wind
density profile and a medium with uniform density, due in part to the
lack of early observations (particularly in the radio and
submillimeter bands) and the degeneracy between dust extinction and
the intrinsic spectral slope in the optical and near-IR bands.  Thus,
the signature of stellar mass loss remains the elusive missing link in
the association of GRBs and massive stars.

To date, the single exception to this disappointing trend is
GRB\,011121, which provides strong evidence for a circumburst
medium shaped by a stellar wind \citep{pbr+02}, and an accompanying SN
\citep{bkp+02,gsw+02}.  The reason for these unambiguous
results is the combination of exquisite {\it Hubble Space Telescope}
(HST) observations, extensive near-IR data, and dual-band radio data. 

More recently, \citet{pkb+02} presented the $\gamma$-ray properties
and redshift ($z=0.695\pm 0.005$) of \grb{}, along with multi-band
ground-based and HST optical observations.  The observations 
between 15 and 65 days after the burst reveal a red bump (with a
spectrum $F_\nu\propto \nu^{-3.9}$) brighter than an extrapolation of
the early data.  \citet{pkb+02} interpret this emission as coming from
a SN accompanying the burst, but note that the statistical
significance of this result depends on the degree of collimation of
the GRB ejecta.  This is because a more collimated outflow results in
an earlier steepening of the afterglow lightcurves, and hence a more
significant deviation at late time.

In this paper we present radio observations of \grb, which point to a
uniform density circumburst medium.  We also show that the radio,
optical, and X-ray data require an early jet break, which
significantly strengthens the SN interpretation for the late-time
emission.  Combining these two results we conclude that a $\rho\propto
r^{-2}$ density profile is not a required signature of a massive
stellar progenitor.

\section{Radio Observations}
\label{sec:obs}

Very Large Array (VLA\footnotemark\footnotetext{The VLA is operated by
the National Radio Astronomy Observatory, a facility of the
National Science Foundation operated under cooperative agreement by
Associated Universities, Inc.}) observations were initiated 1.2 days
after the burst using the standard continuum mode
with $2\times 50$ MHz contiguous bands.  A log of all observations is
given in Table~\ref{tab:rad}.  We used the extra-galactic source
3C\,286 (J1331+305) for flux calibration, while the phase was
monitored using J1356$-$343.  The data were
reduced and analyzed using the Astronomical Image Processing System
\citep{fom81}.

\section{Reverse Shock Emission in the Radio Band}
\label{sec:rad}

\begin{figure}
\plotone{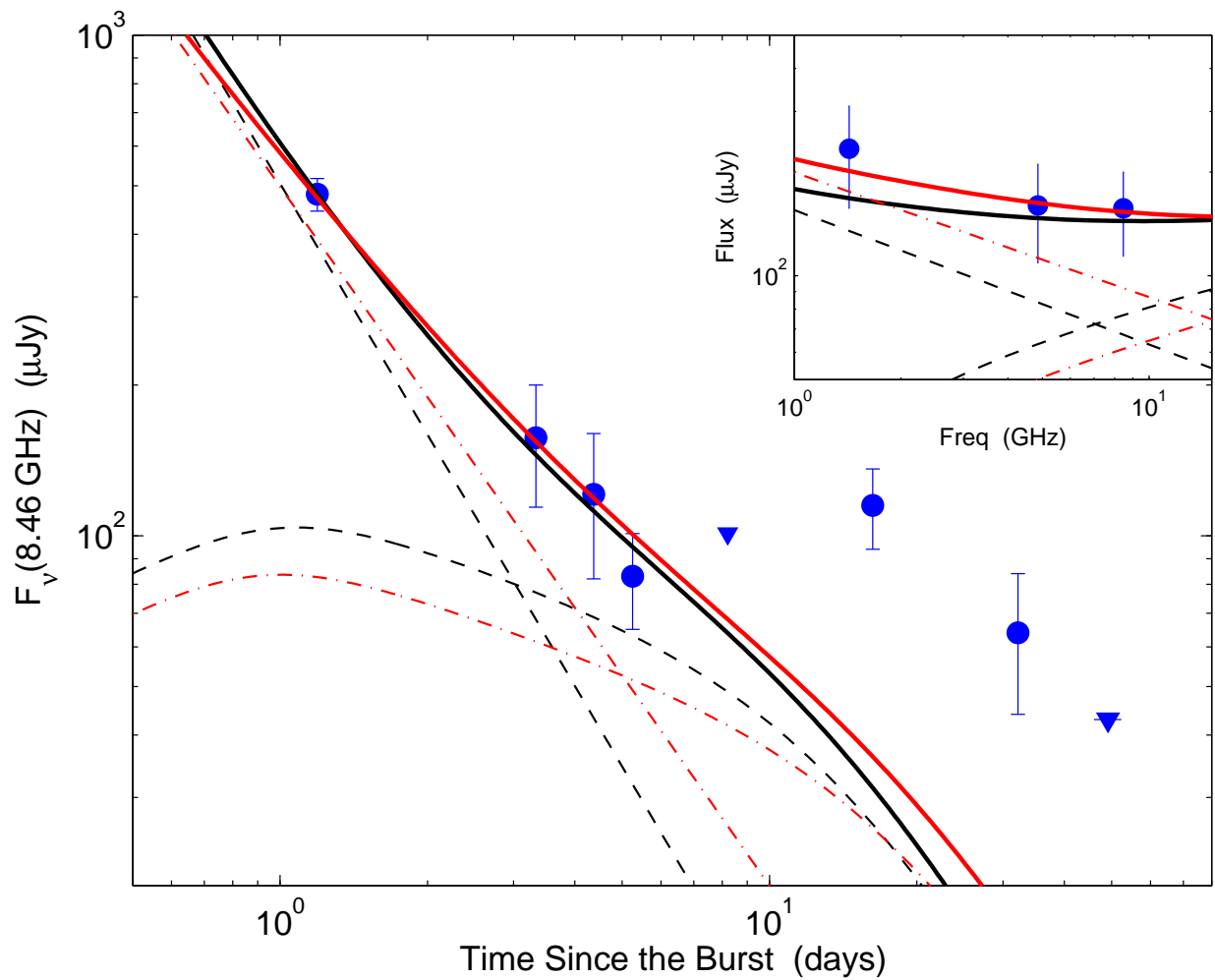}
\caption{Radio lightcurve at 8.46, and the spectrum between 1.4 and
8.5 GHz on day 3.3 (inset).  The solid lines represent the best-fit
combined emission from the reverse and forward shock model described
in \S\ref{sec:models} for the thin and thick shell cases.  The dashed
(thin shell) and dash-dotted (thick shell) lines show the
contributions of the reverse and forward shocks separately.  
\label{fig:rad}}  
\end{figure}

In Figure~\ref{fig:rad} we plot the 8.46 GHz lightcurve of \grb{}, as 
well as the radio spectrum between 1.43 and 8.46 GHz on day 3.3.  The
early emission is characterized by two important features.  First, it
is brightest ($F_\nu\approx 0.5$ mJy) during the first observation
($t\approx 1.2$ days), and rapidly fades, $F_\nu\propto t^{-1.2\pm
0.4}$, between $1.2$ and $5$ days.  Second, the spectral index between
1.43 and 8.46 at $t\approx 3.3$ days is $\beta_{\rm rad}\approx
-0.3\pm 0.3$, and similarly at $t\approx 1.2$ days $\beta_{\rm
rad}<-0.5$ based on the 8.46 and 22.5 GHz data.  

The rapid fading and negative spectral slope are atypical for emission
from the forward shock on the timescale of 1 day.  In fact, for typical
parameters, the radio band, $\nu_{\rm rad}$, lies well below the
peak of the synchrotron spectrum at early time, $\nu_m\approx 260
\epsilon_{B,-2}^{1/2}\epsilon_{e,-1}^2 E_{52}^{1/2}t_d^{-3/2}$ GHz
\citep{se01}; here $\epsilon_{B}=0.01\epsilon_{B,-2}$ and
$\epsilon_{e}=0.1\epsilon_{e,-1}$ are the fractions of shock energy
carried by the magnetic fields and electrons, respectively, and
$E=10^{52}E_{52}$ is the afterglow kinetic energy.  In this regime,
the spectrum is $F_\nu\propto \nu^{1/3}$ (or $F_\nu\propto \nu^2$ if
$\nu_{\rm rad}<\nu_a$, the synchrotron self-absorption frequency;
\citealt{spn98}).  As a result, the early radio flux is expected to
either increase, as $t^{1/2}$ or $t^1$ \citep{spn98,cl00}, be flat
\citep{cl00}, or decay as $t^{-1/3}$ \citep{sph99}, followed by a
steep decline as $t^{-1}$ to $t^{-2}$ when $\nu_m$ crosses the radio
band.  This behavior is observed in the most radio afterglows
(e.g.~\citealt{bsf+00}).  Since the early radio emission from \grb{}
does not follow this general trend, we conclude that it did not arise
from the forward shock. 

Instead, we interpret the observed emission as coming from a reverse
shock \citep{mr97,sp99b} plowing back into the relativisitic
ejecta.  Similar  ``radio flares'' have been observed in several
afterglows \citep{kfs+99,fbg+00,dfk+01}, in particular from
GRB\,990123 where the flare was shown to be the low-energy tail of the
$V\approx 9$ mag optical flash observed at $t\sim 50$ s
\citep{kfs+99,sp99a}.  In all cases in which a radio flare has been
observed, the emission had similar properties to that from \grb{},
namely a bright ($F_\nu\sim 0.5$ mJy) flux measured at early time
($t\sim 1$ day) followed by a rapid decline.  

Hydrodynamical studies of the reverse shock (e.g.~\citealt{ks00}) have
shown that the radio emission depends on the properties of the ejecta
and circumburst medium.  In particular, the main implication of the
radio flare from \grb{} is that it effectively rules out a circumburst
medium with a density profile $\rho\propto r^{-2}$ (hereafter, Wind).
\citet{cl00} have shown that in a Wind environment, with typical
afterglow parameters, the cooling frequency of the reverse shock,
$\nu_{c,{\rm RS}}\sim 4\times 10^{8}t_{\rm sec}$ Hz, is significantly
lower than its characteristic frequency, $\nu_{m,{\rm RS}}\sim 10^{19}
t_{\rm sec}^{-1}$ Hz.  Thus, the emission peaks at $\nu_{c,{\rm RS}}$,
with a flux, $F_{\nu,p}\sim 16$ Jy, independent of time until the
reverse shock crosses the shell at $t_{\rm cr}=5(1+z)\Delta_{10}$ s
\citep{cl00}; here $\Delta=10\Delta_{10}$ light seconds is the initial
width of the shell.  Following the shell crossing, electrons are no
longer accelerated, and since the reverse shock is highly radiative
($\nu_{c,{\rm RS}}\ll \nu_{m,{\rm RS}}$) the emission decays
exponentially.  Thus, strong emission from the reverse shock at $t\sim
1-2$ days is not expected in a Wind environment, indicating that the
early radio emission from \grb{} requires a circumburst medium with
uniform density (hereafter, ISM).  

In addition, based on the flat spectral slope between $1.4$ and $8.5$
GHz measured at $t=3.3$ days, we conclude that for both the reverse
and forward shocks $\nu_a\lesssim 1.4$ GHz.  Otherwise, the emission
from the reverse shock would be severely attenuated over this
frequency range by the forward shock, $F_{\nu,{\rm obs}}=F_{\nu,{\rm
em}}e^{-\tau_\nu}$, resulting in a significantly steeper spectrum;
here $\tau_\nu$ is the synchrotron optical depth.

\section{Uniform Density Models for the Afterglow Emission}
\label{sec:models}

Using the conditions inferred in \S\ref{sec:rad} we model the radio,
optical, and X-ray data with a model describing self-consistently the
time evolution of the forward and reverse shocks in a uniform density
medium.  We consider the optical data only at $t<10$ days since the
emission at later times is dominated by a much redder
component, possibly a SN \citep{pkb+02}.  We return to this point in
\S\ref{sec:prog}.

The time evolution of the reverse shock spectrum ($F_\nu\propto
\nu^{1/3}$ for $\nu<\nu_{m,{\rm RS}}$ and $F_\nu\propto
\nu^{-(p-1)/2}$ for $\nu>\nu_{m,{\rm RS}}$) is described by
$\nu_{m,{\rm RS}}\propto t^{-3(8+5g)/7(1+2g)}$, $F_{\nu,0,{\rm RS}}
\propto t^{-(12+11g)/7(1+2g)}$, and the time of peak emission,
$t_p={\rm max}[t_{\rm dur}/(1+z),t_{\rm dec}]$ \citep{ks00}; here
$t_{\rm dur}=60$ s is the duration of \grb{} \citep{pkb+02}, $t_{\rm
dec}=(3E/32\pi \Gamma_0^8n_0m_pc^2)^{1/3}$, $\Gamma_0$ is the initial
Lorentz factor, and $n_0$ is the circumburst density.  The parameter
$3/2\leq g\leq 7/2$ describes the evolution of the reverse shock
Lorentz factor, $\Gamma\propto r^{-g}$, and the limits correspond to
adiabatic expansion ($g=3/2$) and pressure equilibrium between the
forward and reverse shocks ($g=7/2$).  To evaluate $t_p$, $\nu_{m,{\rm
RS}}(t_p)$, and $F_{\nu,0,{\rm RS}}(t_p)$ we use the physical
parameters of the ejecta and circumburst medium as inferred from the
forward shock emission (see below and Table~\ref{tab:ij}), in
conjunction with equations 7--9 of \citet{kob00} for the thick shell
case (i.e.~when the reverse shock is relativistic and effectively
decelerates the shell), and equations 15--17 for the thin shell case
(i.e.~when the reverse shock cannot decelerate the shell effectively).
We set the nominal values of $g=3/2$ and $7/2$ for the thin and thick
shell cases, respectively.  Thus, the only free parameter of the
reverse shock emission is $\Gamma_0$. 

For the forward shock we use the time evolution of the synchrotron
spectrum in the appropriate regime (i.e.~spherical ISM for $t<t_j$:
\citealt{spn98}, and an expanding jet for $t>t_j$: \citealt{sph99});
here $t_j$, the jet break time, is the epoch at which
$\Gamma\sim\theta_j^{-1}$, and $\theta_j$ is the half opening angle of
the jet.  To account for possible extinction within the host galaxy,
$A_V^{\rm host}$, we use the parametric extinction curves of
\citet{ccm89} and \citet{fm88}, along with the interpolation
calculated by \citet{rei01}.

\begin{figure}
\plotone{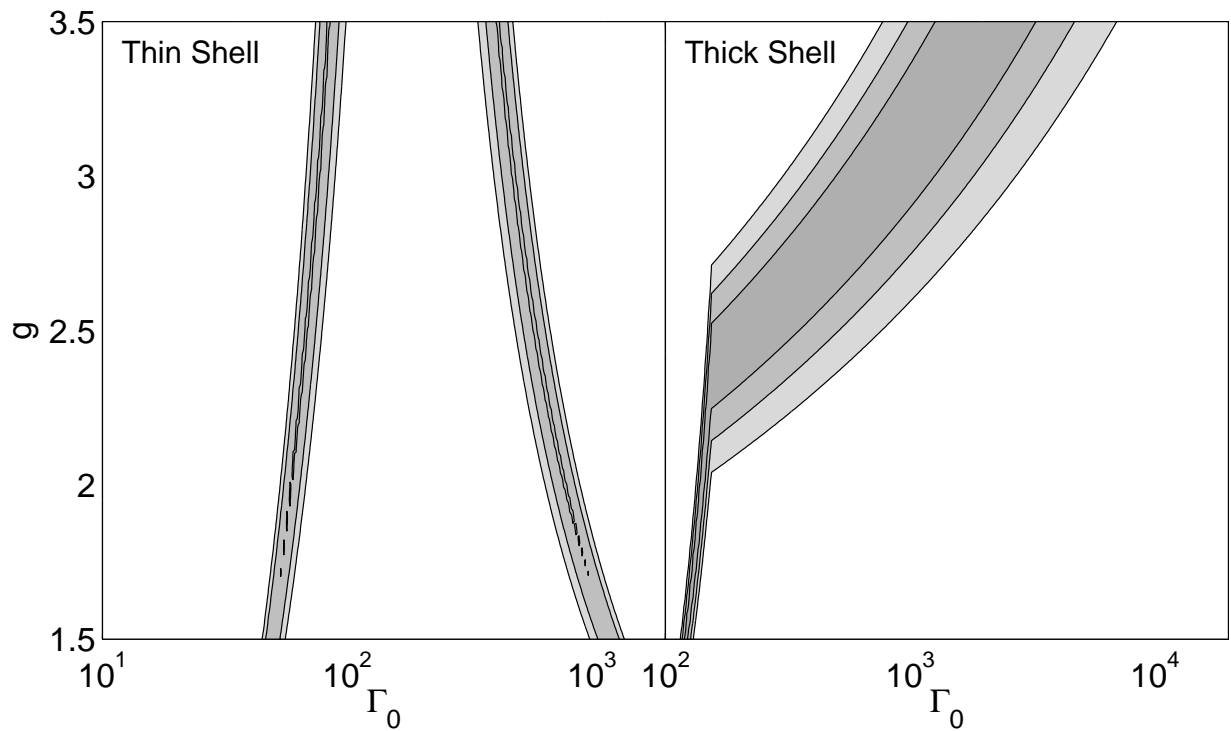}
\caption{Contours (1, 2, and 3 $\sigma$) of $\Gamma_0$ versus $g$
based on the best-fit models for the reverse and forward shock
emission in the thin and
thick shell cases (Table~\ref{tab:ij}).  Both $\Gamma_0$ and $g$ are
not well constrained.  In particular, $\Gamma_0$ has two sets of
minima in the thin shell case, corresponding to the cases when
$t_p=t_{\rm dec}$ and $t_p=t_{\rm dur}/(1+z)$.  The range of
$\Gamma_0$ values in both models encompasses reasonable values based
on expectations from pair production opacity.
\label{fig:cont}}
\end{figure}

The results of the two models (thin and thick shell cases) are shown 
in Figures~\ref{fig:rad}~and~\ref{fig:opt}, and summarized in
Table~\ref{tab:ij}.  We find that both models provide an equally
adequate fit (with $\chi^2_{\rm min}\approx 1$ per degree of freedom),
but $\Gamma_0$ is not well constrained, $\Gamma_0\sim 60-5\times 10^3$
(Figure~\ref{fig:cont}).  

More importantly, both models require collimated ejecta with
$t_j\approx 0.95$ day (hereafter, Jet).  Models with a significantly
wider collimation angle have $\chi^2_{\rm min}\sim 10$ per degree of
freedom, primarily because they underestimate the flux in the X-ray
band by a factor of about $20$, and cannot explain the radio and
optical emission simultaneously.  

At the same time, the Jet models underestimate the radio flux at
$t\gtrsim 15$ days by about $4\sigma$ on day 16.3, and about
$2.5\sigma$ on day 32.  This is due to an apparent brightening of the
radio emission on this timescale.  Since in the radio band
$F_\nu\propto n_0^{1/2}$, one possible explanation for the brightening
is that the forward shock encounters a density enhancement; a density
increase by a factor of ten is required.  The flux in the optical
bands would remain largely unaffected since $\nu_{\rm opt}<\nu_c$ (see
Table~\ref{tab:ij}), in which case the flux is independent of density.

Using the parameters of the forward shock emission we calculate
$E\approx 2\times 10^{51}$ erg, $\epsilon_e\approx 0.1$,
$\epsilon_B\approx 0.5$, and $n_0\approx 10^{-2}$ cm$^{-3}$ using
equations 4.13--4.16 of \citet{se01}.  The opening angle of the jet is
$\theta_j\approx 0.1$ \citep{fks+01}.  Using this value we find a
beaming corrected $\gamma$-ray emergy, $E_\gamma\approx 3.7\times
10^{50}$ erg \citep{pkb+02}, typical for long-duration GRBs
\citep{fks+01}, and a beaming corrected kinetic energy, $E_K\approx
10^{49}$ erg, lower than the typical inferred values of $10^{50}$ to
$3\times 10^{51}$ erg \citep{pk02}.

\section{Implications for the Progenitor of \grb{} and GRB Circumburst
Environments}
\label{sec:prog}

\begin{figure} 
\plotone{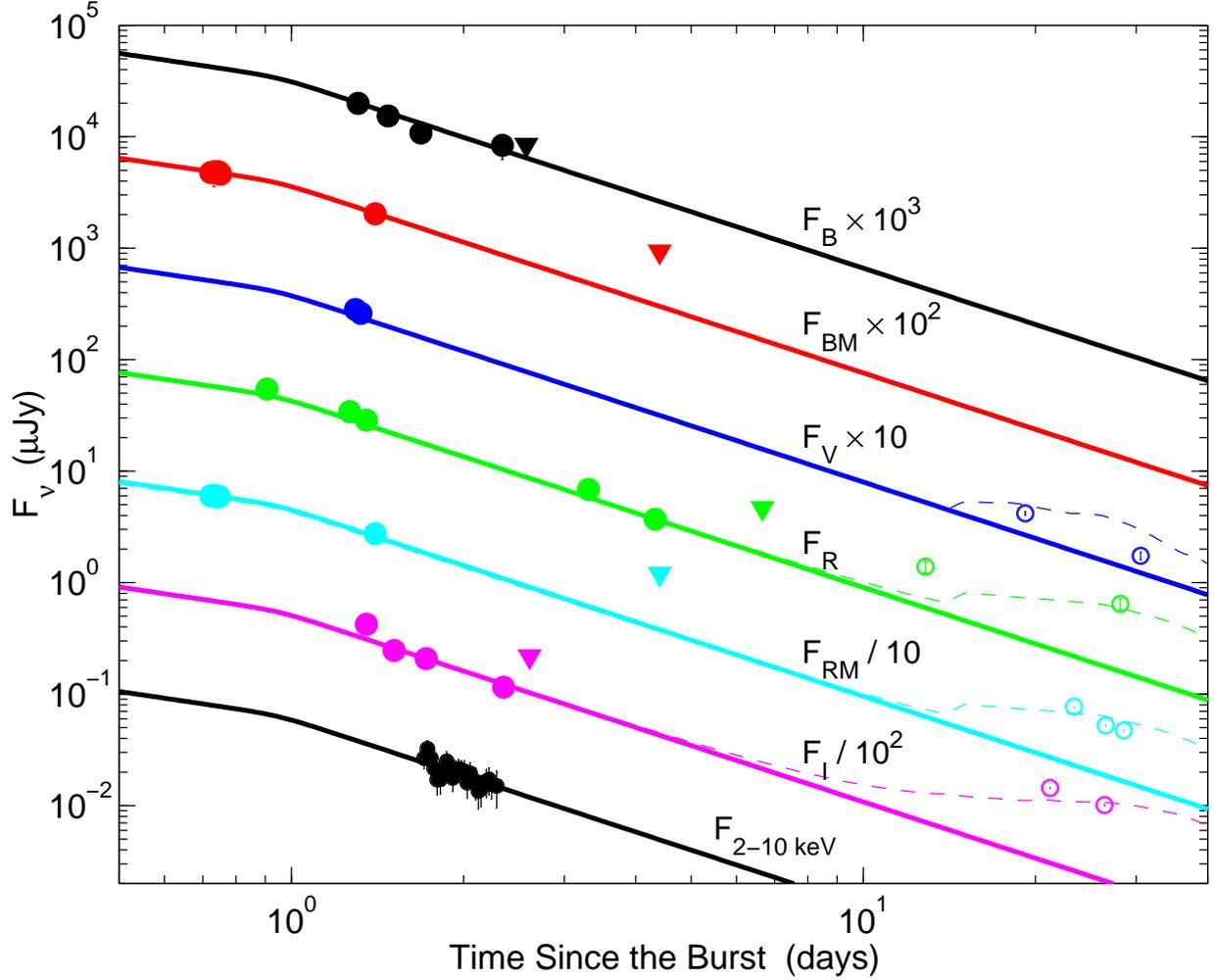}
\caption{Optical and X-ray lightcurves of \grb{}.  The triangles are
upper limits, while the open circles are late-time observations,
dominated by emission from a supernova.  The solid lines represent the
best-fit reverse and forward shock model described in
\S\ref{sec:models}, and the dashed lines are the combined afterglow
flux and flux from SN\,1998bw redshifted to $z=0.695$.  The overall
agreement indicates that the excess emission is due to a SN similar to
SN\,1998bw.
\label{fig:opt}}
\end{figure}

In the previous section we did not consider the optical
emission at $t\gtrsim 10$ days since it has a distinct spectrum,
$F_\nu\propto \beta^{-3.9\pm 0.1}$, compared to $F_\nu\propto
\beta^{-1.15\pm 0.07}$ for the early afterglow data.  Moreover, the
predicted brightness of the afterglow at late time is lower by a
factor of about $3-5$ in the $R$ and $I$ bands compared to the flux
measured with HST (Figure~\ref{fig:opt}).  These two observations
indicate that the late-time emission comes from a separate component.   

\citet{pkb+02} interpreted this emission as coming from a supernova
that occured at about the same time as the burst.  However, they note
that the significance of this conclusion depends sensitively on the
time of the jet break.  Based solely on the optical data, these
authors were unable to significantly constrain $t_j$.  However,
our combined radio, optical, and X-ray model with $t_j\approx 0.95$
day indicates that the SN interpretation is secure. 

We gain further confidence about this interpretation by comparing
the late-time emission to the optical emission from the Type Ic
SN\,1998bw \citep{gvv+98}.  In Figure~\ref{fig:opt} we plot the
combined emission from the afterglow of \grb{} and SN\,1998bw
redshifted to $z=0.695$.  We correct the SN lightcurves for Galactic
extinction, $E(B-V)=0.054$ mag \citep{sfd98}, as well as extinction
within the host galaxy as determined in \S\ref{sec:models}.  While
SN\,1998bw does not provide a perfect fit to the data, the level of
agreement is remarkable given that there are other effects at play
(e.g.~an earlier decay for a fainter luminosity; \citealt{imn+98}).  

The conclusion that \grb{} was accompanied by a supernova indicates
that the progenitor must have been a massive star.  However, we also
demonstrated that the circumburst density profile is uniform, at least
in the range $r\sim 10^{15}$ to $10^{17}$ cm.  This is the first case
in which a massive progenitor and a uniform ambient medium have been 
associated directly, leading us to conclude that a strong density
gradient, $\rho\propto r^{-2}$, on the lengthscales probed by the
afterglow is not a required signature of a massive stellar
progenitor.  In fact, this may explain why the signature of stellar
mass loss has not been observed in the majority of afterglows.

The uniform medium around the progenitor of \grb{} does not imply that
the progenitor did not lose mass.  In fact, it has been previously
suggested that a relatively uniform medium can occur downstream from
the wind termination shock \citep{rdm+01}.  This region can extend out
to $10^{18}$ cm.  In addition, a density enhancement is expected at
the termination shock, which can explain the increased radio flux at
$t\gtrsim 15$ days compared to the Jet model predictions
(\S\ref{sec:models}).

\section{Conclusions} 
\label{sec:conc}

We showed that the early radio emission from \grb{} was dominated by
the reverse shock, and that this directly implies a uniform circumburst 
medium.  The same conclusion holds for all bursts in which radio flares 
have been detected on the timescale of $\sim 1$ day.  The broad-band
data indicate that the ejecta underwent a jet break at $t\approx 0.95$
day (i.e.~$\theta_j\sim 6^\circ$), resulting in a significant
deviation of the late-time optical emission measured with HST from an
extrapolation of the model.  Combined with the spectral properties of
the late emission, and a reasonable agreement with the optical
emission of SN\,1998bw, this indicates that \grb{} was accompanied by
a supernova.  Thus, the progenitor of this burst was a massive star.  

The association of a massive stellar progenitor with a uniform
circumburst medium indicates that the tedious search for a $\rho
\propto r^{-2}$ density profile may have been partly in vain.  It
appears that a Wind profile does not necessarily accompany every GRB,
and in fact may not be the case for the majority of bursts.  This
result, in conjunction with the inferred Wind medium for GRB\,011121
which was also accompanied by a supernova, indicates that the
circumburst environments of GRBs are more diverse than the simple
assumption of constant mass loss rate and wind velocity; the
interaction of the wind with the local environment may play a
significant role.  Deeper insight into the structure of the
circumburst medium requires rapid localizations and dense, multi-band
follow-up.

\acknowledgements 

EB thanks R.~Chevalier and R.~Sari for valuable discussions.  We
acknowledge support from NSF and NASA grants.

\clearpage
\begin{deluxetable}{lllc}
\tabcolsep0.1in\footnotesize
\tablecolumns{4}
\tabcolsep0.1in\footnotesize
\tablewidth{0pc}
\tablecaption{VLA Radio Observations of \grb{} \label{tab:rad}}
\tablehead {
\colhead {Epoch}        &
\colhead {$\Delta t$}   &
\colhead {$\nu_0$}      &
\colhead {Flux Density} \\
\colhead {(UT)}         &
\colhead {(days)}	&
\colhead {(GHz)} 	&
\colhead {($\mu$Jy)}
}
\startdata
Apr 6.22 & 1.19 & 8.46 & $481\pm 36$ \\
Apr 6.25 & 1.22 & 22.5 & $<300$ \\
Apr 8.36 & 3.33 & 8.46 & $157\pm 43$ \\
Apr 8.38 & 3.35 & 4.86 & $160\pm 51$ \\
Apr 8.40 & 3.37 & 1.43 & $234\pm 77$ \\
Apr 8.42 & 3.39 & 22.5 & $<600$ \\
Apr 9.40 & 4.37 & 8.46 & $121\pm 39$ \\
Apr 10.28 & 5.25 & 8.46 & $83\pm 18$ \\
Apr 13.24 & 8.21 & 8.46 & $47\pm 27$ \\
Apr 21.28 & 16.3 & 8.46 & $115\pm 21$ \\
May  7.20 & 32.2 & 8.46 & $64\pm 20$ \\
May 21.12 & 46.1 & 8.46 & $61\pm 26$ \\
May 24.13 & 49.1 & 8.46 & $-9\pm 20$ \\
May 27.17 & 52.1 & 8.46 & $-11\pm 25$ \\\hline
May 21.12--27.17 & 46.1--52.1 & 8.46 & $0\pm 14$
\enddata
\tablecomments{The columns are (left to right), (1) UT date of
each observation, (2) time since the burst, (3) observing frequency,
and (4) flux density at the position of the radio transient with the
rms noise calculated from each image.  The last row gives the flux
density at 8.46 GHz from a co-added map of the data from May 21--27.}  
\end{deluxetable}

\clearpage
\begin{deluxetable}{lcc}
\tablecolumns{5}
\tabcolsep0.1in\footnotesize
\tablewidth{0pc}
\tablecaption{Uniform Density Jet Models for \grb{} \label{tab:ij}} 
\tablehead {
\colhead {Parameter} &
\colhead {Thin Shell}   &
\colhead {Thick Shell}
}
\startdata
$\nu_a$ (Hz)     & {$10^{9}$} & {$10^{9}$} \\
$\nu_m$	(Hz) 	 & {$1.9\times 10^{12}$} & {$4.0\times 10^{12}$} \\
$\nu_c$	(Hz) 	 & {$3.5\times 10^{14}$} & {$1.0\times 10^{14}$} \\
$F_{\rm \nu,0}$ ($\mu$Jy)  & {$740$} & {$760$} \\
$p$ 		 & {$1.78$} & {$1.73$} \\
$t_j$ (days) 	 & {$0.99$} & {$0.91$} \\
$A_V^{\rm host}$ (mag) 	   & {$0.28$} & {$0.25$} \\
$\Gamma_0\,^a$   & {$1.2\times 10^3$} & {$3.2\times 10^3$} \\
$\chi^2_{\rm min}/{\rm dof}$ & {$61.2/61$} & {$58.7/61$} \\ 
$E_{\rm iso,52}$ & $0.3$   & $0.2$  \\
$n_0$		 & $0.05$  & $0.08$  \\
$\epsilon_e$	 & $0.1$   & $0.1$   \\
$\epsilon_B$	 & $0.3$   & $0.7$   \\
\enddata
\tablecomments{Best-fit synchrotron parameters, and the inferred
physical parameters for the reverse and forward shock model described
in \S\ref{sec:models}.  The quoted values for the forward shock are at
$t=t_j$.  We give the results for the thin shell (with $g=3/2$) and
thick shell (with $g=7/2$) cases.  In both models with fix the value
of $\nu_a$ at 1 GHz based on the flat spectrum in the radio band
between 1.4 and 8.5 GHz (\S\ref{sec:rad}).  $^a$ $\Gamma_0$ is not
well constrained and can take a wide range of values 
(Figure~\ref{fig:cont}).}  
\end{deluxetable}

\clearpage

\clearpage

\clearpage

\end{document}